\documentclass[final,3p,times,twocolumn,amsmath,amssymb,amsfonts,multirow]{elsarticle}
\usepackage{graphicx}

\def\Eq#1{Eq.~(\ref{#1})}

\def\0#1#2{\frac{#1}{#2}}

\def\s0#1#2{\mbox{\small{$ \frac{#1}{#2} $}}}

\newcommand{\STr}{\mathrm{STr}}

\newcommand{\be}{\begin{eqnarray}}
\newcommand{\ee}{\end{eqnarray}}

\newcommand{\nn}{\nonumber }
\newcommand{\fslash}{\hspace{-0.1ex} \slash }

\newcommand{\Nc}{N_{\rm{c}}}

\newcommand{\text}[1]{ {\rm #1} }
\journal{Physics Letters B}
\begin{document}
\begin{frontmatter}

\title{On the Phase Structure of QCD in a Finite Volume}
\author{Jens Braun} 
\address{Theoretisch-Physikalisches Institut,
  Friedrich-Schiller-Universit\"at Jena, 07743 Jena, Germany}
\address{TRIUMF, 4004 Wesbrook Mall, Vancouver, BC V6T 2A3,
  Canada} 
  \author{Bertram Klein} 
  \address{Physik Department,
  Technische Universit\"at M\"unchen, James-Franck-Strasse 1, 85748
  Garching, Germany } 
  \author{Bernd-Jochen Schaefer}
\address{Institut f\"ur Physik, Karl-Franzens-Universit\"at, 8010
  Graz, Austria}
  \address{Institut f\"ur Theoretische Physik, Justus-Liebig-Universit\"at Giessen, 35392 Giessen, Germany}

\begin{abstract}
  The chiral phase transition in QCD at finite chemical potential and
  temperature can be characterized for small chemical potential by its
  curvature and the transition temperature. The curvature is
  accessible to QCD lattice simulations, which are always performed at
  finite pion masses and in finite simulation volumes. We
    investigate the effect of a finite volume on the curvature of the
    chiral phase transition line.
    We use functional renormalization group methods with a two flavor
    quark-meson model to obtain the effective action in a
  finite volume, including both quark and meson fluctuation
  effects. Depending on the chosen boundary conditions and the pion
  mass, we find pronounced finite-volume effects. For periodic quark
  boundary conditions in spatial directions, we observe a {\it
    decrease} in the curvature in intermediate volume sizes, which we
  interpret in terms of finite-volume quark effects. Our results have
  implications for the phase structure of QCD in a finite volume,
  where the location of a possible critical endpoint might be shifted
  compared to the infinite-volume case.

\end{abstract}


\end{frontmatter}
%


%
\section{Introduction}
An important theoretical method for the exploration of the QCD phase
diagram at finite temperature and quark chemical potential is the
simulation of the theory on a finite space-time lattice. While this
approach is fully non-perturbative and incorporates most relevant physical
effects, the price to be paid is a restriction to finite simulation
volumes and to pion masses which only recently approach physical
values \cite{Fodor:2004nz}. In addition, the complex phase of the
fermion determinant at finite quark chemical potential makes lattice
simulations with Monte-Carlo methods for finite-density systems
difficult.

Different strategies have been developed to cope with this sign
problem in lattice QCD. 
For example, the partition function can be
evaluated with re-weighting techniques \cite{Fodor:2001au,Fodor:2001pe,Fodor:2002km} 
or Taylor-expanded around vanishing chemical potential \cite{Allton:2002zi,Allton:2003vx,%
Allton:2005gk,Gavai:2003mf,Gavai:2004sd,Gavai:2008zr}. Alternatively,
using an imaginary quark chemical potential for which no sign problem
is present, the theory can be simulated at finite chemical potential,
and the results can be continued to real values of the chemical
potential
\cite{deForcrand:2002ci,deForcrand:2002yi,deForcrand:2003bz}. The
convergence of the Taylor-expansion method has been investigated
e.g. in \cite{Gavai:2008zr} and has recently been studied in an
effective low-energy model \cite{Wagner:2009pm,Karsch:2010hm}. For a
review of the state of lattice techniques see
e.g. \cite{Philipsen:2005mj,Schmidt:2006us,Philipsen:2008gf,deForcrand:2010ys}.

Both approaches have been used to calculate the curvature of the QCD
phase transition line in the temperature and baryon chemical potential
plane, albeit for different values of the number of quark flavors, the
size of the simulation volume and the pion mass
\cite{deForcrand:2002ci, deForcrand:2002yi,deForcrand:2006pv,Karsch:2003va,Kaczmarek:2011zz,Endrodi:2011gv}. At
present, the results from both approaches with different parameter
values differ significantly, and our understanding of these results
should be improved by a better grasp of the influence of these
parameters.

In this context low-energy models of QCD are very useful to
investigate the effects, e.g., of explicit chiral symmetry
breaking or of a
finite volume, and to provide a framework for understanding better the
mechanisms of chiral symmetry breaking. Examples include NJL-type
models and quark-meson-type models, see e.~g. Refs.~\cite{Jungnickel:1996aa,Jungnickel:1995fp,Berges:1997eu,Schaefer:1999em,Braun:2003ii,
  Schaefer:2004en, Braun:2004yk, Braun:2005gy,Braun:2005fj,Abreu:2006pt,Schaefer:2006ds,Stokic:2009uv,Palhares:2009tf,Palhares:2011jf,Fraga:2011hi}, 
  as well as Polyakov-loop extended 
  versions thereof, see e.~g. Refs.~\cite{Meisinger:1995ih,Pisarski:2000eq,Mocsy:2003qw,Fukushima:2003fw,%
  Megias:2004hj,Ratti:2005jh, Roessner:2006xn,Sasaki:2006ww,Schaefer:2007pw,Schaefer:2009ui,Mizher:2010zb,Skokov:2010wb,Herbst:2010rf,Skokov:2010uh}.

For the physics of such models in a finite volume, in addition to the
fermionic fluctuations, the fluctuations of bosonic fields are of
utmost importance~\cite{Braun:2004yk,Braun:2011uq}. 
For a continuous symmetry such 
as the chiral flavor symmetry, bosonic fluctuations of Goldstone modes
restore the symmetry in a finite-volume system in the absence of
explicit symmetry breaking. Consequently, no phase with spontaneously
broken chiral symmetry exists in the chiral limit. In order to capture the effects
of long-range fluctuations, we use a functional renormalization group~(RG) 
approach in the formulation according to
Wetterich~\cite{Wetterich:1992yh}. For reviews of
and introductions to this functional RG approach we refer the reader to
{Refs.~\cite{Litim:1998nf,Bagnuls:2000ae,Berges:2000ew,Polonyi:2001se,Delamotte:2003dw,%
Pawlowski:2005xe,Gies:2006wv,Schaefer:2006sr,Delamotte:2007pf,Rosten:2010vm,Pawlowski:2010ht,Braun:2011pp}.}
This approach can also be adapted to a finite-volume system, see Refs.~\cite{Braun:2004yk, Braun:2005gy, Braun:2005fj,
  Braun:2008sg, Klein:2010tk,Braun:2010vd,Braun:2011uq,Tripolt:2011xx}.

When a system involving fermions is put into a finite Euclidean box,
the anti-commutation relations demand anti-periodic boundary
conditions for the fields in the Euclidean time direction. 
{In contrast, the boundary conditions in spatial directions are not
determined by a physical constraint, and one is free to choose either
periodic or anti-periodic boundary conditions~\cite{Carpenter:1984dd}.}
In lattice QCD
simulations, often spatial periodic boundary conditions for the quark
fields are chosen to minimize finite-volume effects. For small volume
size, the presence of a resulting spatial zero-momentum mode for the
quark fields can affect the formation of the chiral condensate. The
choice of boundary condition leads to significant differences in the
results for the condensate and the Goldstone spectrum
\cite{Braun:2005gy} as well as for the transition temperature
\cite{Braun:2005fj}. Contrary to naive expectations, contributions
from the zero-mode can enhance the chiral condensate in a finite
volume, while for either choice of boundary condition chiral symmetry
is restored in the small-volume limit and ultimately the condensate
vanishes \cite{Braun:2005gy}. A similar effect in the pion mass has
been observed in Dyson-Schwinger equations~\cite{Luecker:2009bs}.

As a consequence of this observation, we expect that the phase
transition line at finite quark chemical potential and temperature
will be affected by a finite volume. In NJL-type model calculations,
the chiral condensate gives mass to constituent quarks, and an
enhancement of the chiral condensate leads to a corresponding increase
in the constituent quark mass. At finite quark chemical potential,
more massive quarks lead to a decrease of the sensitivity of the
system to changes in the quark chemical potential. Hence, we expect
that the chiral transition temperature is less sensitive to the
chemical potential and the curvature of the transition line in the
$\mu$-$T$--plane decreases for periodic spatial quark boundary
conditions in those volume ranges where this effect pertains. 
{In a recent {contribution to proceedings}~\cite{Klein:2010tk}, we have briefly reported on first results 
from an investigation of this hypothesis for values of the volume and the pion mass
that are relevant for current lattice simulations. In this letter, we detail our
studies on this question and present new results for the physical value of the 
pion mass. In addition, we employ a regularization scheme different
from the one used in Ref.~\cite{Klein:2010tk}. This constitutes a nontrivial check
of our earlier results. As we shall discuss below, our new results {do} indeed confirm
that the curvature of the chiral phase transition line 
has an intriguing dependence on the volume size.
}

In Sec.~\ref{sec:RGFV} we introduce the RG equations in a finite
volume for a quark-meson model truncation. The results for the
curvature of the phase-transition line at finite temperature and
chemical potential are presented and discussed in
Sec.~\ref{sec:results}. We summarize our results, draw 
conclusions and present an outlook in
Sec.~\ref{sec:conclusions}.

\section{Renormalization Group Equations in a Finite Volume}\label{sec:RGFV}
We use for our investigation the quark-meson model in a finite volume. At the ultraviolet (UV) 
scale $\Lambda$, the model is defined by the bare effective action
\be
&& \hspace*{-1cm} \Gamma_{\Lambda}[\bar q,q,\phi]= \int d^{4}x \Big\{
 \frac{1}{2} Z_{\phi}(\partial_{\mu}\phi)^{2}+U_{\Lambda}(\phi^2) - c \sigma \Big\} \nn\\
&&\qquad  +\bar{q} \left(Z_{\psi}{\rm i}{\partial}\!\!\!\slash + 
    {\rm i}g(\sigma+i\vec{\tau}\cdot\vec{\pi}\gamma_{5}) + {\rm i} \gamma_0 \mu\right)q
\label{eq:QM}
\ee
with $\phi^{\mathrm{{T}}}=(\sigma,\vec{\pi})$.  The mesonic potential
at the UV scale is parameterized by two couplings, $m^2_\Lambda$ and
$\lambda_\Lambda$,
\begin{equation}
  \label{eq:pot_UV} 
  U_\Lambda(\phi^{2}) =
  \frac{1}{2}m_\Lambda^{2}\phi^{2} +
  \frac{1}{4}\lambda_\Lambda(\phi^{2})^{2}
  \,.
\end{equation}
A current quark mass term $m_c \bar{q} q$ which explicitly breaks the
chiral symmetry has been bosonized and leads to a term $-c \sigma$
linear in the radial $\sigma$ field. The symmetry-breaking parameter
$c$ is related to the quark mass through a combination of the UV
parameters. 

In this work we study the RG flow of the effective action in leading order 
of the derivative expansion (LODE)\footnote{This approximation is also known as the local potential approximation~(LPA).},
where a (possible) space dependence of the expectation value of the scalar
fields is not taken into account and the wave-function renormalizations~$Z_{\phi}$ and~$Z_{\psi}$
are considered to be constant,~$Z_{\phi}\equiv 1$ and~$Z_{\psi}\equiv 1$. 
At finite temperature, this implies that we also neglect a possible difference of the wave-function renormalizations
parallel and perpendicular to the heat-bath~\cite{Braun:2009si}. 
These approximations should by no means be confused with a mean-field (large-$\Nc$) approximation.
On the contrary, the LODE already includes effects beyond the mean-field limit. 
A detailed discussion of the relation of the present approximation to the mean-field approximation in 
terms of a derivative expansion and a large-$\Nc$ expansion of the effective action
can be found in Refs.~\cite{Braun:2008pi,Braun:2009si,Braun:2010tt}. In studies of the quark-meson model, 
however, the anomalous dimensions associated with~$Z_{\phi}$ and~$Z_{\psi}$ are found to be small~\cite{Berges:1997eu}. 
Therefore we believe that the LODE is already sufficient for this
study.

\begin{table*}[t!]
\begin{center}
\begin{tabular}{|c|c|c|c|c|}
\hline 
$m_\pi^{(0)}$ [MeV] & $f_\pi^{(0)}$ [MeV] & $m_\Lambda$ [GeV]  & \phantom{0} $\lambda_\Lambda$ \phantom{0} & $c$ [GeV$^3$] \\
\hline
\hline 
$100$& $90$ & $1.003$  & $100$ & $9.02\times 10^{-4}$ 
\tabularnewline
\hline 
$138$ & $92$ & $0.998$ & $100$ & $1.76\times 10^{-3}$
\tabularnewline
\hline 
$200$ & $97$ & $0.982$  & $100$ & $3.88\times 10^{-3}$  
\tabularnewline
\hline 
\end{tabular}
\end{center}
\caption{Initial UV values for the RG calculation with $N_{\rm max}=2$ for the three values of the 
pion mass considered in the main results. Here, $m_{\pi}^{(0)}$ 
is the pion mass in infinite volume for vanishing temperature and chemical
potential, whereas~$f_{\pi}^{(0)}$ denotes the respective value for the pion decay constant. 
We have fixed the parameters such that the values for $m_{\pi}^{(0)}$ 
and~$f_{\pi}^{(0)}$ are consistent with chiral perturbation theory~\cite{Colangelo:2003hf}. 
All results have been obtained with a UV cutoff $\Lambda = 1.5~\text{GeV}$ 
and a constant Yukawa coupling $g= 3.258$.
\label{tab:UVvalues}}  
\end{table*}

For our derivation of the RG flow equations for the quark-meson model
in a finite volume we use the \mbox{Wetterich} equation~\cite{Wetterich:1992yh}
\begin{equation} 
  \label{eq:Flow_EffAction} 
  \partial_t \Gamma_{k} = \frac{1}{2} \STr\left\{  \left[\Gamma _{k}^{(2)} + R_k\right]^{-1} \left(\partial_t
      R_k\right) \right\} \ .
\end{equation}
For explicit calculations, we employ optimized regulator 
functions, see Refs.~\cite{Litim:2006ag, Blaizot:2006rj}. 
For details on optimization of RG flows, we refer the reader to 
Refs.~\cite{Litim:2000ci, Litim:2001up,Litim:2001fd,Pawlowski:2005xe}.
At finite temperature and in a finite volume, the $3d$ regulator allows us to separate the
finite-volume contributions from the finite-temperature contributions,
which yields an exceptionally simple form for the RG flow equation.
To be specific, we employ the following regulator functions for
bosonic (B) and fermionic (F)  fields:
\begin{eqnarray} 
R_{\mathrm{B}} (p_0,\vec{p}\,) &=&\vec{p}^{\,2} r_{\mathrm{B}} (\vec{p}^{\,2}/k^2)\quad\text{and} \nonumber \\
R_{\mathrm{F}} (p_0,\vec{p}\,) &=&  \fslash\hspace{-1.2ex}\vec{p}\,r_{\mathrm{F}}(\vec{p}^{\,2}/k^2)\ .
\end{eqnarray}
The shape functions $r_{\mathrm{B}} (x)$ and $r_{\mathrm{F}} (x)$ are given
explicitly by
\begin{eqnarray} 
  r_{\mathrm{B}} (x)&=&\left(\frac{1}{x} - 1\right)\Theta (1-x)\quad \text{and}\nonumber \\
  r_{\mathrm{F}} (x)&=&\left(\frac{1}{\sqrt{x}}-1\right)\Theta (1-x)\ .
\end{eqnarray}

In order to derive the RG flow equations for a system in a finite
four-dimensional Euclidean volume $L^{3}\times 1/T$ with temperature
$T$, we replace each spatial momentum integral in the evaluation of
the trace in \Eq{eq:Flow_EffAction} by a sum over discrete momenta:
\begin{equation} 
  \int_{-\infty}^\infty dp_{i} \rightarrow
  \frac{2\pi}{L} \sum_{n_{i}  \in \mathbb{Z}}\quad;\quad i=1,2,3\ . 
\end{equation} 
While the boundary conditions in the Euclidean time direction are fixed by the statistics
of the fields, we are free in the choice of the boundary conditions for the bosons and fermions in
the spatial directions. In the
following we use the short-hand notation 
\begin{equation}
\vec{p}^{\,2}_{\text{p}}=\frac{4\pi^{2}}{L^{2}}\sum_{i=1}^{3}n_{i}^{2}\;\;
\mathrm{and} \;\;
\vec{p}^{\,2}_{\text{a}}=\frac{4\pi^{2}}{L^{2}}
\sum_{i=1}^{3}\left(n_{i}+\frac{1}{2}\right)^{2} 
\label{eq:fv_mom}
\end{equation}
for the three-momenta in the case of periodic (p) and anti-periodic
(a) boundary conditions.  The flow equation for the effective
potential of the quark-meson model in a finite cubic volume with
length $L$ at finite temperature $T$ and quark chemical potential
$\mu$ is then given by
  \begin{eqnarray}
    \label{eq:fv_ft_flow_equation}
   \partial_t U_k(\phi^2)
    &=& k^5 \Bigg[ \frac{3}{E_\pi} 
    \left( \frac{1}{2}+n_B(E_\pi) \right){\mathcal B}_{\text{p}}(kL) \nn \\
    && \quad\quad\quad
    +\frac{1}{E_\sigma} \left(
      \frac{1}{2}+n_B(E_\sigma)\right){\mathcal
      B}_{\text{p}}(kL) \nn\\
   && -\frac{2 N_c N_f}{E_q} \Big( 1-n_F(E_q,\mu) \nn\\ 
   && \quad\quad\quad\;\; - n_F(E_q,-\mu)
    \Big){\mathcal B}_{\text{l}}(kL) \Bigg] \, ,
  \end{eqnarray}
where the first two terms correspond to contributions of the mesonic
modes, and the last term with opposite overall sign corresponds to the
quark contributions, see also Ref.~\cite{Braun:2010vd}.
The effective quasi-particles energies are given
by
\begin{equation}
E_i=\sqrt{k^2+M_i^2} \; , \quad i \in \{ \pi, \sigma, q \} \, , 
\end{equation}
with the corresponding mesonic
\begin{eqnarray} 
  M_{\pi}^2 = 2 \frac{\partial
    U}{\partial \phi^2}\,,&&\quad
  M_{\sigma}^2 = 2 \frac{\partial U}{\partial \phi^2} + 4 \phi^2
  \frac{\partial^2 U}{\partial (\phi^2)^2}
\end{eqnarray} 
and squared  quark masses
\begin{equation} 
  M_q^2 = g^2 \phi^2.\label{eq:constqm}
\end{equation}
The usual bosonic and fermionic occupation numbers read
\begin{equation}
  n_B(E)=\frac{1}{e^{E/T}-1} \; , \;\; n_F(E,\mu)=\frac{1}{e^{
      (E-\mu)/T}+1} \, . 
\end{equation}
The dependence on the finite spatial volumes is encoded in the
mode counting functions ${\mathcal B}_{\text{l}}$: 
\begin{equation} 
  {\mathcal B}_\text{l}(kL)=\frac{1}{(kL)^3} \sum_{\vec{n} \in \mathbb{Z}^3} 
  \Theta\!\left( (kL)^2    
    - \vec{p}_{\text{l}}^{\,2}L^2\right)\,,
\end{equation} 
{where $\text{l}\in\{\text{a},\text{p}\}$ and~$\vec n$ labels the 
three-dimensional vector of integers.}
Depending on the choice for the spatial boundary conditions for the quarks, the 
appropriate function ${\mathcal B}_{\text{l}}$ appears in the last term of~\Eq{eq:fv_ft_flow_equation}.
For small $kL$ we find for periodic boundary conditions 
\begin{eqnarray} 
\lim_{kL\to 0} {\mathcal B}_{\text{p}}(kL) \sim \frac{1}{(kL)^3} 
\end{eqnarray}
and for antiperiodic boundary conditions
\begin{eqnarray}
\lim_{kL\to 0} {\mathcal B}_{\text{a}}(kL) = 0\ . 
\end{eqnarray} 
The behavior of ${\mathcal B}_{\text{p}}$ for periodic spatial
boundary conditions reflects the fact that the dynamics of the system
for small $kL$ is mainly governed by the spatial zero modes.

{{The} flow equation~(\ref{eq:fv_ft_flow_equation}) for the chiral
order-parameter potential can be coupled straightforwardly to the 
confinement order-parameter potential, namely the Polyakov-loop potential.
{We give the flow equation including the dependence on the minimum of the Polyakov-loop potential in~\ref{app:pol}}. 
In the following, however, we will not consider such an extension
in our numerical analysis but restrict ourselves to the flow equation (\ref{eq:fv_ft_flow_equation}). 
{The possible consequences of this restriction are discussed in the conclusion, together with the implications of our results.}
}

In the limit of infinite volume ($L\to \infty$) we find for both
boundary conditions, $\text{l}\in\{\mathrm{a},\mathrm{p}\}$,
\begin{equation}
  \lim_{kL\to\infty} {\mathcal B}_{\text{l}}(kL) = \frac{1}{6\pi^2}\ .
\end{equation} 
As expected, in this limit the same flow equation for the effective
potential for infinite volume is recovered which has been found with a
similar regulator function for $L\to\infty$ in \cite{Schaefer:2004en,
  Braun:2003ii, Stokic:2009uv}.

In order to solve the RG flow for the scale-dependent effective 
mesonic potential $U_k$, we expand the potential in a Taylor series in
scale dependent local $n$-point couplings $a_{n, k}$ around its scale-dependent 
minimum $\sigma_{0, k}$ 
\begin{eqnarray}
  \label{eq:pot_ansatz}
  U_k(\phi^2) &=&  \sum_{n=0}^{N_{\text{max}}}\frac{a_{n,k}}{2^n n!}
  ( \phi^2\!-\! 
  \sigma_{0,k}^2)^{n}.
\end{eqnarray} 
Due to the presence of the symmetry-breaking term $-c\sigma$, the
minimum is shifted from its value in the chiral limit.  The condition
\begin{equation} 
  \label{eq:min_cond}
  \frac{\partial}{\partial
    \sigma} U_k(\sigma^2+\vec{\pi}^{\,2}) \Bigg|_{\vec{\pi}=\vec{0},\sigma=\sigma_{0,k}} \stackrel{!}{=} c
\end{equation} 
ensures that the potential is always expanded around the actual
physical minimum \cite{Braun:2008sg}. From \Eq{eq:min_cond} one sees that the RG flow of
the coupling $a_{2, k} \equiv m_k^2$ and the minimum $\sigma_{0, k}$ are
related by the simple condition
\begin{equation}
  \label{eq:min_cond2}
  a_{2,k} \sigma_{0,k} = c\,. 
\end{equation} 
This condition keeps the potential minimum at $(\sigma, \vec{\pi}) =
(\sigma_{0,k}, \vec{0}\,)$.

The RG flow equations for the couplings $a_{n,k}$ and $\sigma_{0, k}$
can be obtained by expanding the equation for the effective potential,
\Eq{eq:fv_ft_flow_equation}, around the running minimum $\sigma_{0,
  k}$ and then projecting it onto the derivative of the ansatz, \Eq{eq:pot_ansatz}, with respect to $k$. 
  In general, this procedure results
in an infinite set of flow equations for all $a_{n,k}$. In order to
obtain a finite set of flow equations, we truncate the Taylor series,
\Eq{eq:pot_ansatz}, at a fixed order $N_{\text{max}}$ and
  include thus fluctuations around the minimum up to order
$2N_{\text{max}}$ in the mesonic fields.  The convergence of such an
expansion in powers of the fields, i.e., in $n$-point functions, has
been studied quantitatively in Ref.~\cite{Tetradis:1993ts} at
vanishing temperature by computing critical exponents and
for the LODE at finite temperature in Ref.~\cite{Papp:1999he}.
The resulting set of coupled first-order differential equations can
then be solved numerically for example by standard Runge-Kutta
methods.

{In the following we use~$N_{\rm max}=2$ and fix the {para\-meters}~$m_\Lambda^2\equiv a_{1,\Lambda}$, 
$\lambda_\Lambda\equiv a_{2,\Lambda}/2$,~$c$
and~$g$ such that, at zero temperature and chemical potential,
 the values of the pion mass  and the pion decay constant are consistent 
 with chiral perturbation theory~\cite{Colangelo:2003hf}, see Tab.~\ref{tab:UVvalues}. 
While we consider the couplings~$m_\Lambda$ and~$\lambda_\Lambda$ to
be RG-scale dependent, we keep the Yukawa coupling constant in the present study,~$g=3.258$.
Together with the value of the pion decay constant, the Yukawa coupling then determines the constituent
quark mass, see~\Eq{eq:constqm}. Note that the four-boson coupling is marginal.
For the UV cutoff~$\Lambda$ (initial RG scale), we choose $\Lambda = 1.5~\text{GeV}$. 
{This} scale can be {interpreted} as a hadronic mass scale below which 
hadronic operators are considered to be the relevant degrees of freedom
and the dynamics {are} dominated by light pions. Of course, the precise values
of the UV cutoff and the couplings {are} scheme-dependent. However, our main
results are consistent with results that have been obtained using a different scheme, 
see Ref.~\cite{Klein:2010tk} and {the} discussion below, provided that the parameters are chosen such that
the values of the IR observables are identical.
}

\begin{table*}[t!]
\begin{center}
\begin{tabular}{|c|c|c|c|c|c|}
\hline 
method & ref. & $N_f$  & $a m_c$ &  $L^3 \times 1/T$& $\kappa$
\tabularnewline
\hline
\hline 
 imaginary $\mu$ & \cite{deForcrand:2002ci} & $2$ & $0.032$ & $8^3\times 4$ & $0.500(54)$ 
\tabularnewline
\hline 
 imaginary $\mu$ & \cite{deForcrand:2006pv} & $3$ & $0.026$ & $8^3 \times 4$ & $0.667(6)\phantom{4}$ 
\tabularnewline
\hline 
 Taylor series& \cite{Karsch:2003va} & $3$ & $0.005$ & $12^3 \times 4$, $16^3 \times 4$ & $1.13(45)\phantom{4}$ 
\tabularnewline
\hline 
Taylor series& \cite{Kaczmarek:2011zz} & $2+1$  & \phantom{000}$^{\it a}$ & $32^3 \times 8$ & \phantom{1}$0.58(2)(4)$
\tabularnewline
\hline
Taylor series& \cite{Endrodi:2011gv} & $2+1$ & \phantom{000}$^{\it a}$ 
& $24^3\times 8, 28^3\times 10$ &\phantom{1}$0.59(18)^{\it b}$\phantom{1}
\tabularnewline
\hline 
\end{tabular}
\end{center}
{\small
\vspace*{-0.3cm}
\hspace*{3cm}${}^{\it a}$ quark masses close to physical values, more than one value\\\noindent
\hspace*{3cm}${}^{\it b}$ result from chiral condensate\phantom{xxxxxxxxxxxxxxxxxxxxxxxx}}
\caption{Results for the curvature $\kappa$ from lattice simulations
  for $N_f=2$, $N_f=2+1$ and $N_f=3$ flavors, for different values of the current
  quark mass $a m_c$ (in dimensionless lattice units) and 
  volume sizes $ L^3 \times 1/T$ (in units of the lattice spacing
  $a$). \label{tab:lattcurv}}  
\end{table*}

\section{The phase boundary in a finite volume}\label{sec:results}
In the following we discuss finite-volume effects on the shape of the
chiral QCD phase diagram at small chemical potentials.  A very useful
quantity for a description of the phase transition close to the
temperature axis is the curvature $\kappa$ of the transition line at
$\mu=0$, since it is accessible to both QCD lattice simulations and
phenomenological models. It appears as the first non-vanishing
coefficient in a Taylor series expansion of the transition line
$T_\chi(\mu)$ in powers of $\mu^2$ around vanishing chemical potential. It
is defined according to
\be
  &&\hspace*{-1cm}\frac{T_{\chi}(\mu, L,m_{\pi}^{(0)})}{T_{\chi}(\mu\!=\!0,L,m_{\pi}^{(0)})} \nn\\
  &&\quad =1-\kappa(L, m_\pi^{(0)})
  \frac{\mu^2}{\pi^2 T^2_{\chi}(0,L,m_{\pi}^{(0)})} + \dots\,,
\ee
where $m_{\pi}^{(0)}=m_{\pi}(T=0, \mu\to 0,L\to\infty)$ is the pion
mass in infinite volume for vanishing temperature and chemical
potential. Thus, the volume dependence of the phase diagram for small
chemical potentials is encoded in the curvature $\kappa$ which depends
on $L$ and $m_{\pi}^{(0)}$.

In Table~\ref{tab:lattcurv}, we list some results for the curvature
$\kappa$ obtained with different lattice methods. Only the result for
$N_f=2$ flavors from \cite{deForcrand:2002ci} could be compared
directly to our calculations. The comparison between the different
lattice results \cite{Karsch:2003va,deForcrand:2006pv} for $N_f=3$
shows that the values differ significantly with a change of both the
lattice current quark mass $am_c$ and the simulation volume $L^3
\times 1/T$.  We investigate the effect of the volume on the curvature
to see if such an effect contributes to these differences.

Before we present our results for the volume dependence of $\kappa$,
we briefly discuss its dependence on $m_{\pi}^{(0)}$ in infinite
volume ($L\to\infty$). 

In the chiral limit, results for the curvature
in infinite volume have been obtained in Ref.~\cite{Schaefer:2004en}  for the same model considered here.
For a Taylor expansion of the effective potential, as in the present paper, 
to order $\phi^4$ ($N_{\rm max} =2$), a value $\kappa = 1.27$  has been found.
This value is compatible with the results from the present calculation.
In comparison, solving for the full effective potential on a grid, a value of $\kappa = 1.14$ has been
obtained \cite{Schaefer:2004en}.

For finite pion masses, we show our results
for the curvature in infinite volume
$\kappa(L\to\infty,m_{\pi}^{(0)})$ in Tab.~\ref{tab:curv_inf} as a function of the pion mass
$m_{\pi}^{(0)}$.  We observe that $\kappa$ increases slightly with
increasing pion mass. This behavior is in contrast to general
expectations~\cite{Toublan:2005rq} and (lattice) gauge theory studies, where it was found that $\kappa$
decreases with increasing $m_{\pi}$, i.e., increasing current quark
mass $m_c$. This can be understood in terms of the constituent quark
picture: The constituent quark mass $m_q$ increases for increasing
current quark mass $m_c$.  Hence, the system becomes less sensitive to
the chemical potential at any given value.  Since the chiral phase
temperature has only a weak dependence on $m_c$, the curvature $\kappa
\sim T_\chi(0) {d}^2T_\chi(\mu)/ {d}\mu^2$ becomes
smaller. While the first part of the argument holds for the
quark-meson model as well, the chiral phase transition temperature has
a comparatively strong dependence on $m_c$ in model
studies~\cite{Berges:1997eu, Schaefer:1999em, Schaefer:2004en,
  Braun:2005fj}. This is in contrast to lattice studies, see e.~g. Ref.~\cite{Karsch:2000kv}.
   Because the expansion parameter
$\mu^2/T^2_{\chi}(0,L, m_{\pi}^{(0)})$ in our definition of the curvature
is normalized with the transition temperature, this strong dependence
also enters the coefficient $\kappa$.  From the combination of both
effects, we find that the resulting coefficient for the curvature
increases slightly with increasing $m_c$.

\begin{figure}
\includegraphics[clip=true,scale=0.30]{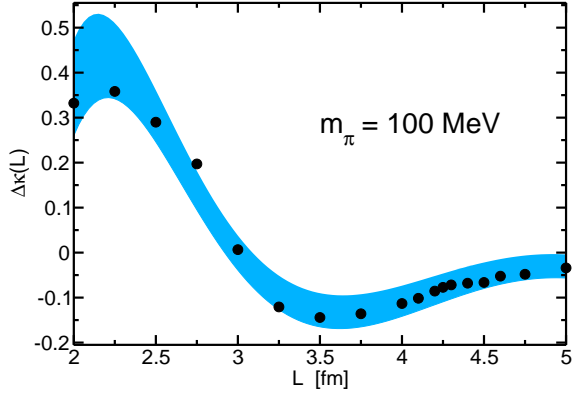}
\includegraphics[clip,scale=0.30]{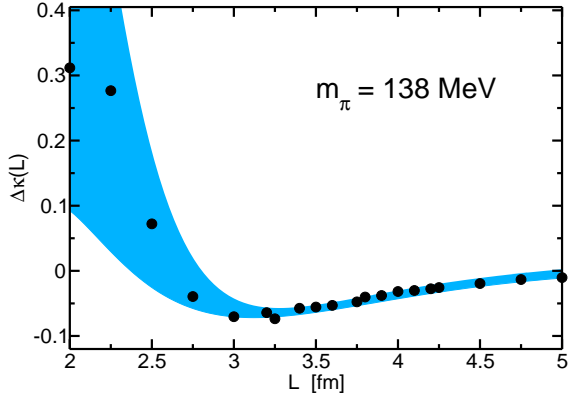}
\includegraphics[clip,scale=0.30]{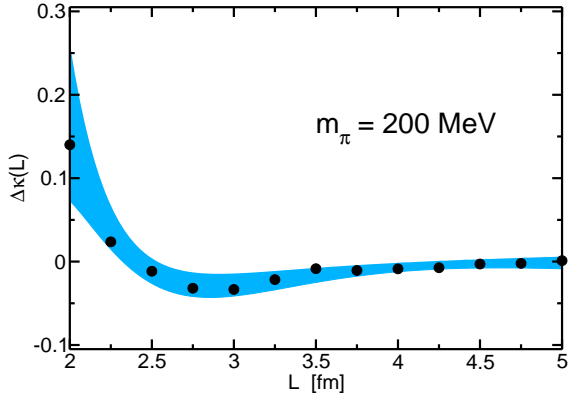}
\caption{$\Delta\kappa(L)=({\kappa(L,m_{\pi}^{(0)})-\kappa(\infty,m_{\pi}^{(0)})})/
  {\kappa(\infty,m_{\pi}^{(0)})}$
  as a function of the spatial extent $L$ of the system for various
  values of $m_{\pi}^{(0)}$. In intermediate volume ranges the curvature is reduced in 
  comparison to its infinite-volume value. For small pion masses, the change can be as much
   as $20 \%$; for a realistic value of the pion mass it is on the order of $10 \%$ and 
   decreases further with increasing pion mass. Error bands are estimated from different 
   fit orders ($\mu^2$, $\mu^4$, $\mu^6$) to the curve $T_\chi(\mu)$.}
\label{fig:curv} 
\end{figure}

\begin{figure}[t]
\centering\includegraphics[clip,scale=0.30]{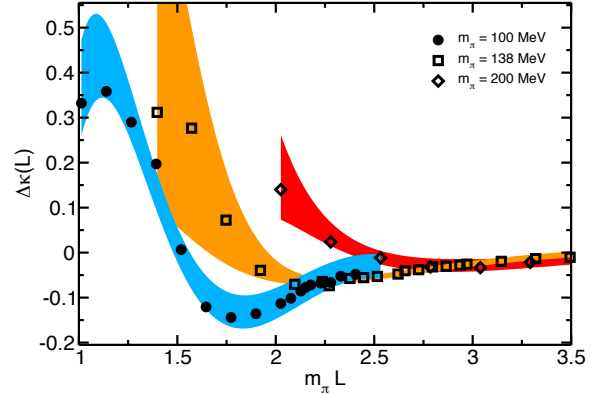}
\caption{$\Delta\kappa(L)=({\kappa(L,m_{\pi}^{(0)})-\kappa(\infty,m_{\pi}^{(0)})})/
  {\kappa(\infty,m_{\pi}^{(0)})}$
  as a function of the dimensionless variable $m_\pi^{(0)} L$ of the system for all three
  values of $m_{\pi}^{(0)}$ investigated.}
\label{fig:curvmpiL} 
\end{figure}

{In Figs.~\ref{fig:curv} and~\ref{fig:curvmpiL} we show our results for the relative change in
the curvature $\Delta\kappa= (\kappa(L,m_{\pi}^{(0)})-
\kappa(\infty,m_{\pi}^{(0)}))/ {\kappa(\infty,m_{\pi}^{(0)})}$ as a
function of the volume size $L$ and of the dimensionless 
quantity~$m_{\pi}^{(0)}L$ for three different values of the pion {mass $m_{\pi}^{(0)}$.}}
For a given value of~$L$, the black {symbols} represent the median 
values for~$\Delta\kappa$ which have been obtained from
fits to the transition curve $T_\chi(\mu)$ with polynomials to 
order~$\mu^2$, $\mu^4$ and~$\mu^6$.
The error bands for $\Delta \kappa$ (shaded areas) are 
also estimated from these fits.
In addition, systematic truncation errors may be present.
In the large volume limit, we find that $\kappa$
approaches its infinite-volume limit as expected.  On the other hand,
we find that the curvature increases rapidly for small volume
sizes. This can be understood from the fact that shrinking the spatial
volume of the system has a similar effect as increasing the
temperature of the system which corresponds to shrinking the extent of
the system in Euclidean time direction. Therefore the constituent
quark mass decreases eventually with decreasing $L$. In turn, the
system becomes more sensitive to the presence of a finite chemical
potential and the curvature $\kappa$ becomes bigger.

Between these two asymptotical limits of large and small volume sizes,
the shape of the curvature $\kappa (L)$ as a function of the volume
size depends on the choice of the boundary conditions for the quarks
in spatial directions as well as on $m_{\pi}^{(0)}$. Since we are
interested in a comparison with lattice simulations, where generally
periodic boundary conditions are chosen, we will concentrate on this
case in the following.  In particular, we can define a critical volume
size $L_c$ below which $\kappa(L)$ increases, and we find that its
value depends on $m_{\pi}^{(0)}$. We observe that $L_c$ becomes
smaller for larger values of $m_{\pi}^{(0)}$.
In Tab.~\ref{tab:Lcmpi} we list the values of the length scale $L_{\rm
  c}$, for which we take the location of the minimum in $\Delta
\kappa$, for selected values of the pion mass $m_\pi(L \to \infty, T
\to 0)$.
In the IR regime, the value of the pion mass in infinite volume and
for zero temperature provides the relevant scale. When the spatial
extent of the system becomes comparable to the Compton wavelength of
the pion, $L\sim 1/m_{\pi}^{(0)}$, the dynamics of the system become
affected by the finite size of the volume. 
{For $ m_{\pi}^{(0)} L \ll 2\pi$, the dynamics}
of the theory are completely dominated by the zero
modes of the fields and the curvature of the phase boundary exceeds
its value in the infinite volume limit. In addition, we observe that
$\kappa(L)$ has a minimum, which appears only in the case with
periodic boundary conditions for the quarks in spatial directions.

The existence of such a minimum is closely related to the existence of
a maximum in the constituent quark mass as a function of $L$ for
periodic boundary conditions~\cite{Braun:2005gy}.  In a constituent
quark picture, the increase of the quark mass leads to a less chemical
potential sensitivity of the system and hence to a decrease in the
curvature. Here this manifests itself as a clear finite-volume effect.

We find that the minimum of the curvature becomes deeper for a
decreasing pion mass $m_{\pi}^{(0)}$ and for a fixed system size
$L$. For example, the curvature deviates about $2\%$ at
its minimum at $L \simeq 3\,\text{fm}$ from its infinite-volume limit for
$m_{\pi}^{(0)}=200\,\text{MeV}$, while it deviates at the minimum almost $20\%$
for $m_{\pi}^{(0)}=100\,\text{MeV}$. For
$m_{\pi}^{(0)}\gtrsim 300\,\text{MeV}$, the deviation of the curvature
from its infinite volume limit is less than $1\%$ for $L\gtrsim
2\,\text{fm}$. These observations imply that an extrapolation to large
volumes is rather simple and safe for large pion masses,
$\kappa(L)\approx \kappa(\infty)$ for $L\gtrsim
2\,\text{fm}$. However, for physical pion masses, the presence of the
minimum in the curvature probably requires more elaborate
extrapolation methods.  In this case finite-volume effects are much
more pronounced, which may have consequences for lattice QCD studies.
\begin{table}[t!]
\begin{tabular}{|c||c|c|c|}
\hline 
$m_{\pi}\,\mathrm{[{MeV}]}$&
100&
138&
200
\tabularnewline
\hline
\hline 
$\kappa(L\to\infty)$&
1.357(18)&
1.375(63)&
1.409(59)
\tabularnewline
\hline 
\end{tabular}
\caption{Dependence of the curvature $\kappa(L\to\infty)$ on the pion mass 
$m_{\pi}(T=0,L\to\infty)$. {The {errors} 
result from fits of the numerical data to the transition curve with polynomials to 
order $\mu^2$ , $\mu^4$ and $\mu^6$.}\label{tab:curv_inf}
}
\end{table}

{We would like to point out that a second extremum exists for (very) small box sizes. In our {results, this} is most clearly
visible for $m_{\pi}^{(0)}=100\,\text{MeV}$, but the onset of the flattening of~$\Delta \kappa$ associated with such 
an extremum is also visible for $m_{\pi}^{(0)}=138\,\text{MeV}$. 
{The region in which significant effects can be described by our model is limited.}
First of all, we {observe} that the values of {the momenta of the} 
non-zero (spatial) momentum modes increase for decreasing volume size as $1/L$. 
For sufficiently small volumes, we then
have $2\pi/L > \Lambda$, where~$\Lambda$ is the UV cutoff.
{We} expect that this small-volume regime is not accessible within our present model approach;
in the same way as the high-temperature phase of QCD cannot be described accurately {with our model} due
to the lack of gauge degrees of freedom {at short length scales} . Since the behavior of the curvature for (very) small volumes may very well be influenced
by the fact that our model is not valid on all scales, we have to restrict ourselves to the regime with~$L\gtrsim 2\pi/\Lambda\simeq 1\,\text{fm}$. 
Second, the length scale set by the pion mass needs to be compared to the extent of the volume. 
For $L\lesssim 2\pi/m_{\pi}^{(0)}$, the
size of the box is smaller than the Compton wavelength of the pion and the partition function of the theory
is dominated by the zero modes, as discussed above. 
Since the pion mass is the relevant scale for the low-energy dynamics,
the positions of the extrema of~$\Delta \kappa$ can be changed by varying the pion mass~$m_{\pi}^{(0)}$. To be specific,
the extrema are shifted to larger values of~$L$ when we decrease the pion mass~$m_{\pi}^{(0)}$.
{Eventually,} the constituent quark mass becomes {for all practical purposes} independent of~$L$ for small volume sizes, $L\ll 2\pi/m_{\pi}^{(0)}$,
provided that we choose periodic boundary conditions for the quarks, {as discussed in} 
Ref.~\cite{Braun:2005gy}. Therefore the chiral phase transition temperature~\cite{Braun:2005fj} 
as well as the curvature remain finite and become {almost} independent of~$L$ in this 
regime.\footnote{{The} constituent quark mass tends to zero monotonically when we use antiperiodic 
boundary conditions in the spatial directions for the quarks~\cite{Braun:2005gy}. In this case, the quarks do not have
a spatial zero mode and the curvature tends to zero for small volumes sizes, $1/L \gg \mu$, see Ref.~\cite{Tripolt:2011xx}.}
}

\begin{figure}[t]
\centering\includegraphics[clip,scale=0.75]{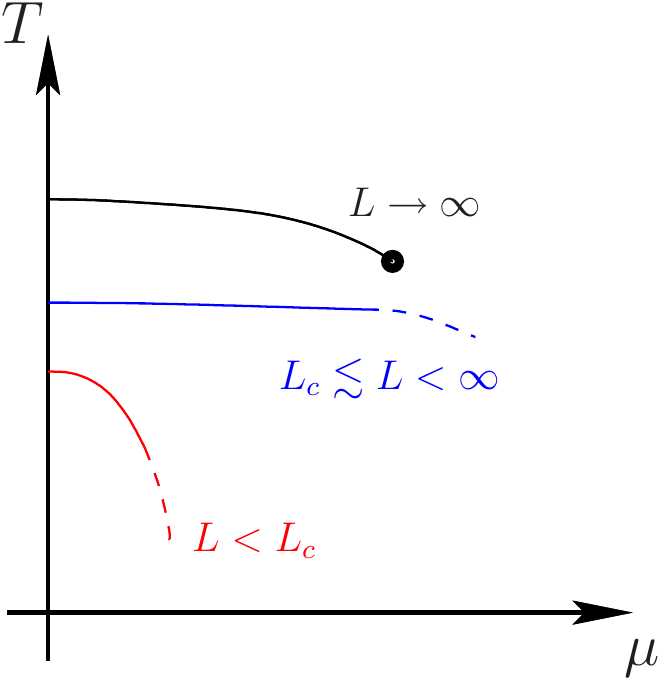}
\caption{Sketch of the QCD phase diagram for small chemical potentials
  for different volume sizes $L$. The straight lines symbolize the
  chiral crossover line. The black dot symbolizes the critical
  endpoint of the chiral phase diagram from a quark-meson model
  study, e.g.~\cite{Schaefer:2004en}.}
\label{fig:pd_sketch} 
\end{figure}

In Fig.~\ref{fig:pd_sketch} we show a sketch of the chiral phase
diagram for different finite volumes as obtained from our quark-meson
model study.  The black line denotes the crossover line in the
infinite-volume limit. The black dot symbolizes the critical endpoint
of the phase diagram as found in
e.g.~Ref.~\cite{Schaefer:2004en}. From our model study we expect that
the chiral QCD phase boundary flattens for $L_c\lesssim L < \infty$.
Therefore the location of a possible critical endpoint might very well
be shifted to larger or smaller values of the chemical potential. For
$ m_{\pi}^{(0)} L \gg 2\pi$, the chiral phase diagram shrinks
drastically both in the temperature direction as well as in the
$\mu$-direction. This means that the location of a potentially
existing critical endpoint is necessarily located at small chemical
potentials, see red line in Fig.~\ref{fig:pd_sketch}. Since the
critical length scale $L_c$ also depends on the pion mass, this
picture is overall dependent on the value of $m_\pi$. 

Our present results are consistent with
previous results from a proper-time renormalization group approach (PTRG), see Ref.~\cite{Klein:2010tk}. 
For~$L\to\infty$, the regulator in the latter study can be directly related to the regulator
employed in the present work \cite{Litim:2001hk}. In the case of a finite volume, however, 
a one-to-one mapping between the PTRG flow equation and the present functional RG flow equation
is not known. The qualitative agreement of our present study with the results in 
Ref.~\cite{Klein:2010tk} suggests that the general behavior of the curvature as a function
of the volume size is independent  of the regularization scheme.
\section{Conclusions and Outlook}\label{sec:conclusions}
{We have used a non-perturbative functional renormalization group approach}
to investigate how the curvature of the chiral phase transition line
close to $\mu=0$ is affected by the volume size. We employed a
quark-meson model with two flavors at finite temperature and chemical
potential in a finite volume. 
{The various parameters of this model have been fixed such that the values of
the pion mass and the pion decay constant at vanishing temperature and chemical
potential are consistent with chiral perturbation theory.}

\begin{table}[t!]
\begin{tabular}{|c||c|c|c|}
\hline 
$m_{\pi}\,\mathrm{[{MeV}]}$&
$100$&
$138$&
$200$
\tabularnewline
\hline
\hline 
$L_{\rm c}$ [fm]&
$\phantom{-}3.64(2)$& 
$\phantom{-}3.19(8)\phantom{0}$&
$\phantom{-}2.97(22)\phantom{0}$
\tabularnewline
\hline
$m_\pi^{(0)} L_{\rm c}$&
$\phantom{-}1.84(1)$&
$\phantom{-}2.23(6)\phantom{0}$&
$\phantom{-}3.01(22)\phantom{0}$
\tabularnewline
\hline 
$\Delta\kappa(L_{\rm c})$&
$-0.14(2) $&
$-0.067(5) $&
$-0.025(13)$
\tabularnewline
\hline
\end{tabular}
\caption{Critical length scale  $L_{\rm c}$ at which the curvature has a minimum. Results are given 
for three different values of the infinite-volume pion mass 
$m_{\pi}(T=0,L\to\infty)$. To find the minimum, one needs to explore the region where the 
dimensionless quantity $m_\pi^{(0)} L_{\rm c} \approx 2 - 3$.
{The {errors} result from fits of the numerical data to the transition curve with polynomials to 
order $\mu^2$ , $\mu^4$ and $\mu^6$.}
\label{tab:Lcmpi} }
\end{table}

The curvature is accessible to lattice QCD calculations in spite of
the fermion sign problem at non-vanishing chemical potential.  Lattice
results for the curvature obtained for different values of the quark
mass and in different volume sizes differ significantly, and it
appears worthwhile to investigate volume effects.

We find a qualitatively clear picture for the behavior of the
curvature in a finite volume. With periodic boundary conditions for
the quark fields in the spatial directions, as done in many lattice
calculations, the curvature decreases in intermediate volumes, and the
observed curvature is smaller than in the infinite-volume limit. Below
a certain critical length scale, which depends on the value of the
infinite-volume pion mass, the curvature again increases strongly.

We interpret these results in terms of a constituent quark picture: An
increasing constituent quark mass makes the system less sensitive to a
change in the chemical potential and leads to a decrease in the
curvature. For periodic spatial boundary conditions for the quark
fields, such an effect has been observed in RG quark-meson model
studies.  For small volume sizes, similar to the behavior at finite
temperature, chiral symmetry tends to be restored and the constituent
quark mass decreases rapidly which finally leads to a strong increase
in the curvature.

While the pion mass dependence of the chiral transition temperature,
obtained from the model, is stronger than the one observed in
calculations including gauge degrees of freedom, the pion mass
dependence of the observed change in the curvature is more closely
related to the infrared (long-range) physics of the system and hence
likely less dependent on microscopic details of the model.

{In {the} present study, we have found that the dependence of the curvature on the volume size 
is weak for larger pion masses. {Our} results suggest that this dependence 
becomes increasingly larger when realistic pion masses are approached in QCD lattice
simulations. 
The observed effect would lead to a flattening of the chiral phase
transition line in the QCD phase diagram for small values of $\mu$ in
an intermediate volume range.} As a consequence, in this volume range
the location of a possible critical endpoint could well be
shifted. For small volumes, the region with broken chiral symmetry
shrinks dramatically, and a possible critical endpoint would be
shifted to small values of the chemical potential. Studies in this
direction are ongoing and will be published elsewhere~\cite{Tripolt:2011xx}.

{It is interesting to speculate whether the confining dynamics in the gauge sector
{alter} the present results qualitatively. This is of course an important question in gauging 
the validity of the current results and relevant to {making contact} with lattice {simulations
on} a quantitative level. In order to better understand how 
confining dynamics change our results for the curvature, one may
include the Polyakov loop in the present study along the lines of 
Refs.~\cite{Braun:2009gm,Braun:2011fw} {and as shown in~\ref{app:pol}}.
The associated corrections leave the zero-temperature dynamics unchanged. At finite temperature,
however, the quarks are effectively screened due to their coupling to the confinement
order parameter. We expect that the inclusion of this special type of gauge dynamics 
will yield only a quantitative change of
the volume dependence of the curvature. The qualitative behavior of
the curvature as a function of the box size {likely persists}.  A
detailed study of this conjecture also requires the inclusion of finite-volume effects in the gauge sector (see, e.~g., 
Refs.~\cite{Bazavov:2007zz,Fischer:2007pf,Berg:2011ts}) and
is therefore deferred to future work.
}

{Finally, we would like {to} emphasize that our present study is primarily meant to {show} that
the chiral QCD phase boundary may have an intriguing dependence on the volume size.
At the present {stage, a} quantitative comparison of the absolute values for the curvature
with the ones found in lattice simulations is difficult. In fact, the absolute value of the curvature
is not the heart of the present study, but rather an interest in the change of the curvature
when the volume of the system is varied. In this respect, our investigation shows that there is a 
qualitative effect which can lead to smaller curvatures~$\kappa$ in a finite volume when compared
to larger or infinite volumes. This could also partly account for
differences in the curvature observed in QCD lattice simulations in
differently sized volumes and help to guide future lattice studies of this quantity.
}

{\it -- Acknowledgments --} 
The authors are very grateful to H.~Gies and J.~M.~Pawlowski for useful discussions. 
JB acknowledges financial support by the DFG under Grant BR~\mbox{4005/2-1} 
and the DFG research training group GRK~1523/1. 
Moreover, this work was partly 
supported by the Natural Sciences and Engineering
Research Council of Canada~(NSERC). TRIUMF receives federal funding
via a contribution agreement through the National Research Council of
Canada. BK acknowledges support of the DFG research cluster "Structure and Origin of the Universe". 
In addition, this work was supported by the Helmholtz-University Young Investigator 
Grant \mbox{No.~VH-NG-332}.

\appendix

{
\section{Chiral Order-Parameter Potential and the Polyakov Loop}\label{app:pol}

For illustration purposes, we show the flow equation for the effective (order-parameter) potential~$U$
in the limit of vanishing quark chemical potential. It is a straightforward generalization of~\Eq{eq:fv_ft_flow_equation} 
and reads 
\be
\partial_t U_k(\phi^2)
    &=& k^5 \Bigg[ \frac{3}{E_\pi} 
    \left( \frac{1}{2}+n_B(E_\pi) \right){\mathcal B}_{\text{p}}(kL) \nn \\
    && \quad\quad\quad
    +\frac{1}{E_\sigma} \left(
      \frac{1}{2}+n_B(E_\sigma)\right){\mathcal
      B}_{\text{p}}(kL) \nn\\
&& -\frac{2 N_f}{E_q}\sum_{j=1}^{\Nc} 
   \Big( 1-n_F(E_q, - {\rm i}\bar{g} \langle A_0\rangle_j) \nn\\ 
&&\qquad\quad
    - n_F(E_q,{\rm i} \bar{g}\langle A_0\rangle_j)
    \Big){\mathcal B}_{\text{l}}(kL) \Bigg]  \,,\nn
\ee
where~$\langle A_0\rangle$ represents the minimum of the
Polyakov-loop potential, i.~e. the order-parameter potential for
confinement~\cite{Braun:2007bx,Marhauser:2008fz,Braun:2010cy}.  The sum runs over the
eigenvalues of the gluonic background field~$\langle A_0^{a}\rangle T^{a}$.
The~$T^{a}$ denote the generators of the underlying~SU($\Nc$)
gauge group in the fundamental representation. Of course, this flow
equation can be generalized to finite quark chemical potential and can 
also be rewritten such that it depends directly on the
Polyakov-loop $\langle {{\rm tr}_{\rm F}} L[A_0]\rangle = \langle {\rm tr}_{\rm
  F}{\mathcal P} \exp ({\rm i}\bar{g}\int_0^{1/T}dx_0 A_0) \rangle /\Nc$ by
assuming~\cite{Meisinger:1995ih} that~$\langle {\rm tr}_{\rm F} L[A_0]\rangle = {\rm tr_F}L[\langle A_0\rangle]$ as well 
as~$N_{\rm c}^n {\rm tr}_{\rm F} (L[\langle A_0\rangle ])^n = N_{\rm c} ({\rm tr}_{\rm F} L[\langle A_0\rangle ])^{n}$ 
with $n\in {\mathbb N}$, see Refs.~\cite{Skokov:2010wb,Herbst:2010rf,Skokov:2010uh}.
}

\end{document}